# Towards an Extended VLTI: Turbulence Characterization for Kilometer-Scale Optical Links at Paranal

B. Neichel[a,*], T. Pichon[a], G. Bourdarot[b], D. Perez[c], E. Vera[c], V. Chambouleyron[a], T. Fusco[d,a], C.-T. Héritier[d,a], F. Eisenhauer[b], N. Vedrenne[d], D. Delamoye[a], F. Glasser[e], S. Esposito[f]

[a]Aix Marseille Univ, CNRS, CNES, LAM, 13013 Marseille, France
[b]Max-Planck-Institut für extraterrestrische Physik, Giessenbachstraße 1, 85748 Garching, Germany
[c]Pontificia Universidad Católica de Valparaíso, Chile
[d]DOTA, ONERA, Université Paris Saclay, 91123 Palaiseau, France
[e]Apical Technologies, Marseille, France
[1f]INAF - Osservatorio Astrofisico di Arcetri, Largo E. Fermi 5, 50125 Firenze, Italy
*benoit.neichel@lam.fr

## ABSTRACT

Future extensions of the VLTI aim to push infrared interferometry toward kilometer-scale baselines, opening access to angular resolutions of a few tens of micro-arcseconds. A pragmatic first step would be to couple a new telescope installed on the VISTA platform with the existing VLTI infrastructure, thereby creating a 1.4 km baseline within the Paranal Observatory. Among the candidate transport solutions, optical fiber, evacuated pipes, or open-air propagation, direct free-space beam transmission with active adaptive-optics (AO) pre-compensation, an approach inspired by Free-Space Optical (FSO) communication technology, offers the best combination of spectral flexibility, moderate infrastructure cost, and the ability to preserve spatially resolved field information. We present a preliminary AO dimensioning study for this link, covering deformable-mirror fitting error, photon-noise budget, anisoplanatism, and scintillation, which shows that moderate-order correction (of order 10×10 actuators) can be sufficient, but that the achievable field of view and sensitivity depend critically on the largely unknown vertical and along-path distribution of the turbulence. We therefore propose a dedicated turbulence-monitoring experiment across the VISTA–VLTI line of sight, designed to quantify the strength, spatial distribution, and temporal evolution of refractive-index fluctuations over this 1.4 km horizontal path. The experiment will deploy two compact 30–50 cm telescopes facing each other, together with calibrated light sources, and will implement two complementary diagnostic approaches: a wide-field Shack–Hartmann sensor enabling tomographic reconstruction of the turbulence volume, and an event-based (neuromorphic) camera approach whose microsecond-scale temporal resolution and high dynamic range allow detection of fast and anisotropic turbulence signatures. The dataset will provide the first systematic characterization of horizontal turbulence over a kilometer-scale optical link at Paranal, constituting a critical milestone toward a kilometer-baseline extended VLTI.

## 1. INTRODUCTION

The Very Large Telescope Interferometer (VLTI) at Paranal Observatory has revolutionized high-angular-resolution astronomy in the optical and near-infrared domain. With the GRAVITY+ instrument operating on the four 8 m Unit Telescopes and four 1.8 m Auxiliary Telescopes, baselines of up to 130 m provide milli-arcsecond imaging and 30–100 micro-arcsecond astrometry [1]. These capabilities have enabled landmark results ranging from the detection and spectroscopy of exoplanets, to probing general relativity in the Galactic Center, to resolving line emission in distant quasars.

Yet the analogy with radio interferometry suggests that the next transformative step lies in dramatically extending the baseline. As argued by Bourdarot & Eisenhauer [2], kilometer-scale optical/infrared interferometry (KBI) would deliver angular resolutions of order 10 μas at 2 μm, a factor of ten beyond the current VLTI, enabling direct imaging of exoplanet surfaces, access to circumplanetary disk structures at sub-AU scales, and new probes of black-hole physics from galactic to cosmological distances (Fig. 1). The scientific landscape for KBI has been articulated in dedicated ESO Expanding Horizons white papers [3, 4] and in a community science prospects document [5]: targets range from resolving cloud coverage on brown dwarfs to measuring massive binary black holes in the final parsec.

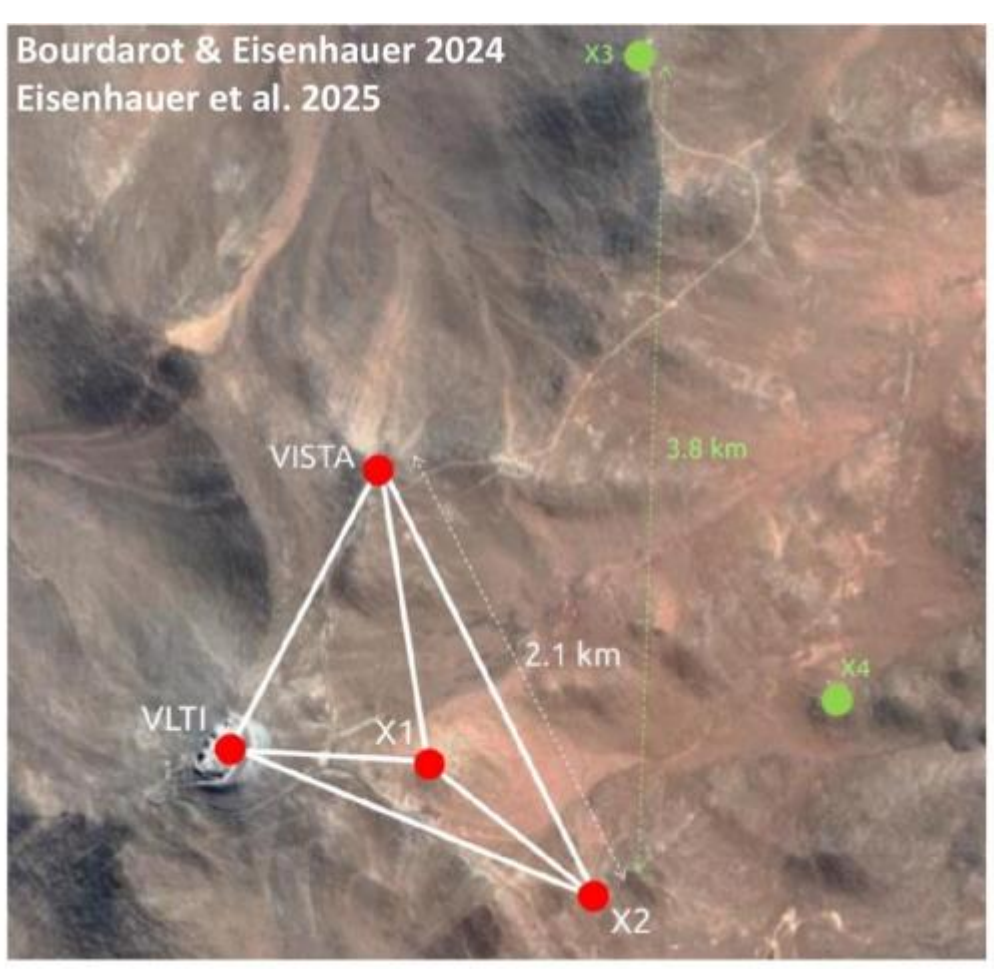

*Figure 1. Sketch of a future Kilometer Baseline Interferometer built around the existing VLTI platform, with outrigger telescopes (X1–X4) reaching baselines of up to 3.8 km. Angular resolution scales as $\Theta = \lambda/2B \approx 10$ μas for B = 10 km (Bourdarot & Eisenhauer 2024; Eisenhauer et al. 2025).*

A natural first step toward this vision is to couple the VISTA survey telescope, located on a ridge 1.4 km from the VLTI platform [6], to the VLTI beam-combining laboratory via a free-space optical (FSO) link. This would immediately provide the longest baseline available at Paranal and serve as a proof-of-concept for open-air beam transport at kilometer scales. The concept is directly motivated by the sensitivity gains achieved with GRAVITY+, which demonstrated that combining extreme adaptive optics (AO), laser guide stars, noiseless near-infrared detectors, and dual-field fringe tracking can overcome the photon-budget limitations that previously constrained ground-based interferometry [1, 2].

The central technical challenge for a VISTA–VLTI FSO link is the characterization and correction of atmospheric turbulence over the 1.4 km horizontal propagation path. Unlike the vertical or near-vertical paths studied in classical telescope AO, a horizontal beam traverses a large fraction of the ground layer and may be affected by thermally driven boundary-layer turbulence, road-induced heating, and other near-ground phenomena whose spatial and temporal structure is poorly known at Paranal. This paper first revisits, in Section 2, why free-space propagation was retained over alternative transport media; Section 3 presents a preliminary AO dimensioning study for the link, quantifying fitting error, photon noise, anisoplanatism, and scintillation; Section 4 describes the experimental concept for the turbulence-characterization campaign; Section 5 details the two complementary sensing approaches and presents prior field-campaign results validating the event-camera methodology; Section 6 presents the expected performance and deliverables; Section 7 concludes.

## 2. SCIENTIFIC AND TECHNICAL CONTEXT

### *2.1 The case for kilometer-baseline interferometry*

Kilometer Baseline Interferometry (KBI) occupies a unique niche in the landscape of future large astronomical facilities. At optical/near-infrared wavelengths, a 10 km baseline delivers an angular resolution of $\theta = \lambda/2B \approx 10$ µas at $\lambda = 2$ µm. Critically, the optical/NIR waveband combines this ultimate resolving power with sensitivity to thermal emission from stars, planets, and gas, access to rich molecular and atomic spectral diagnostics, and transparency of the Earth's atmosphere [5].

The exoplanet science case is particularly compelling. With a 5 km baseline array of six telescopes, one can resolve the surface of Proxima Centauri b, the cloud distribution of nearby brown dwarfs such as Luhman 16, and the circumplanetary disk around young giant planets like β Pic b [3]. Simulations by Bourdarot et al. [3] show that six-telescope, 4 km baseline snapshot imaging of Luhman 16 would reveal large-scale cloud structures at a spatial resolution comparable to present-day Doppler imaging, but without the latitude degeneracy intrinsic to that technique. The planet-formation science case calls for 100 µas imaging resolution and 1 µas astrometry to access circumplanetary disks, resolve terrestrial planet-forming regions, and trace magnetospheric accretion on protoplanets [4]. At the other end of the distance scale, a KBI would reach an angular resolution comparable to that of GRAVITY on the Galactic Center black hole when applied to the supermassive black hole of the Andromeda galaxy (M31) at 765 kpc, opening the diversity of black-hole astrophysics to systematic study.

### *2.2 The VISTA–VLTI baseline as a first milestone*

The VISTA survey telescope is located on an adjacent ridge to the VLTI platform, at a distance of about 1.4 km with a height difference of about 124 m [6]. A terrain study conducted as part of the internship of T. Pichon demonstrated that a direct line of sight exists between the VLTI delay-line tunnel and the VISTA Nasmyth platform, with the beam grazing the terrain at a minimum altitude of approximately 100 m above ground at the lowest point (Fig. 2).

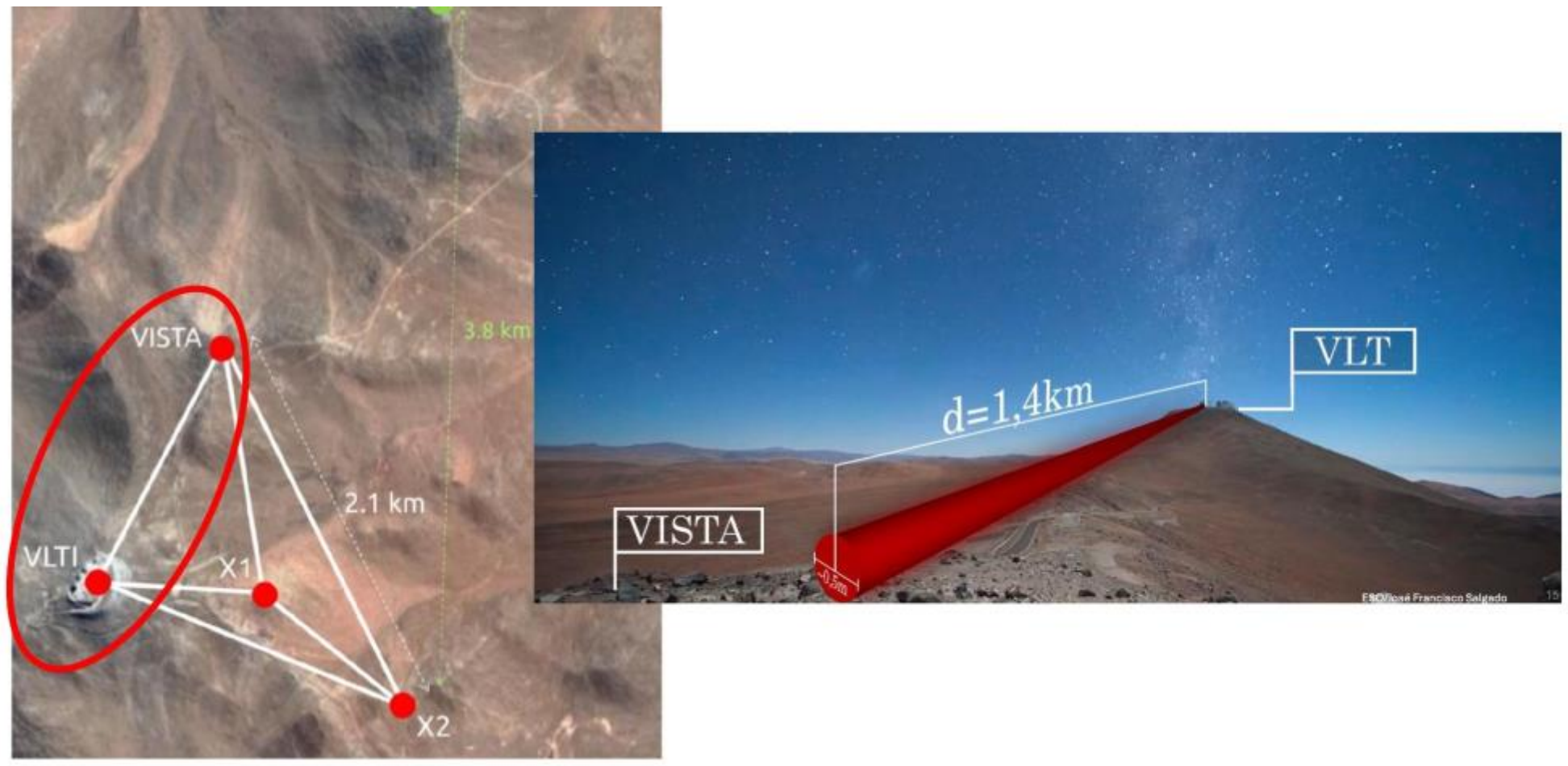

*Figure 2. (a–b) The VISTA–VLTI line of sight across the Paranal ridge. (c) Distance from the beam to the access road along the path. (d) Elevation profile along the line of sight, showing the beam clearance above ground (minimum ≈100 m).*

In this architecture, a 4 m class telescope collects the astronomical source at the VISTA site, while a separate 50–60 cm telescope handles the beam relay across the horizontal path (Section 3).

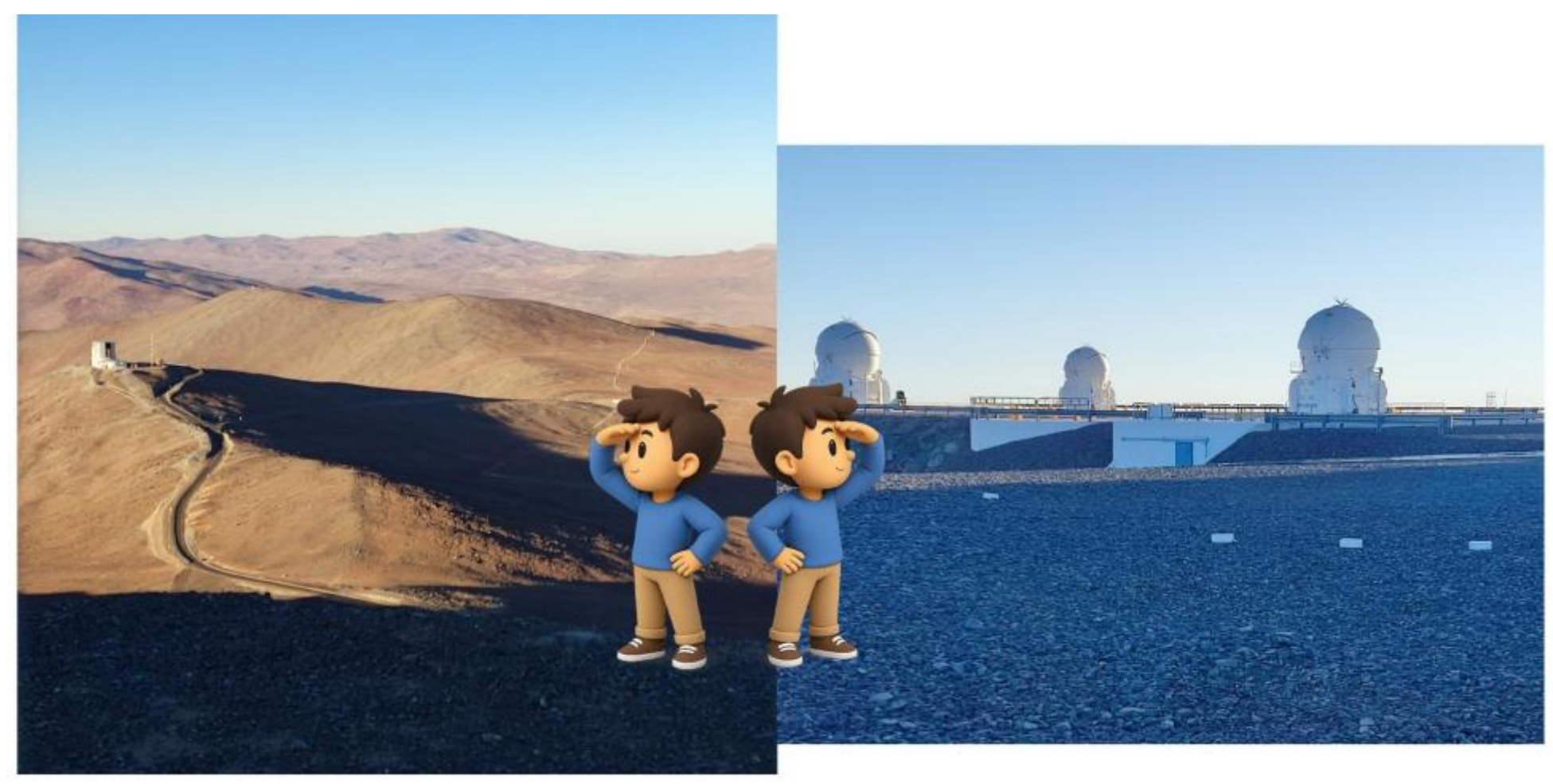

*Figure 3. VISTA–VLTI optical link architecture: a 4 m class telescope collects the science beam at VISTA, which is compressed to a 0.4 m collimated beam by an afocal relay, propagated 1.4 km in free space, and re-imaged onto the VLTI facility pupil by a 50–60 cm receiving telescope.*

## 2.3 Transport options: why free-space propagation

Three families of solutions exist for coupling two telescopes separated by a kilometer-scale distance without digging a new interferometric tunnel: single-mode optical fiber (as pioneered by the 'OHANA project [10]), evacuated light pipes, and open-air free-space transmission with active AO pre-compensation. Table 1 summarizes the trade-off across the criteria most relevant to the VISTA–VLTI case.

*Table 1. Comparison of transport options for the VISTA–VLTI link.*

| Criterion | Optical Fiber ('OHANA, Mariotti+1998) | Vacuum Pipes (Mozurkewich+2016) | Free-Space + AO (Horst+2023) |
|---|---|---|---|
| Topographic constraint | Low (flexible routing) | High (strict line-of-sight) | Moderate (line-of-sight) |
| Spectral coverage | J/H/K bands (silica or fluoride fiber) | All bands (visible → MIR) | All bands (visible → MIR) |
| Turbulence handling | Filtered (single-mode fiber) | None required (vacuum) | AO-corrected (active) |
| Infrastructure cost | Low | High (pipes, pumping) | Low–Moderate |
| Technology readiness | Demonstrated (Keck–Keck fringes) | Conceptual (PFI design study) | Demonstrated in Telecom |
| Main challenge | Attenuation/losses in the K-band fibers. Field transmission | Construction work, earthquakes. | Anisoplanatism for field transmission |

The 'OHANA concept [10] demonstrated that single-mode fibers placed after an AO system can transport an already-corrected beam over hundreds of meters with high fidelity, and remains the most technologically mature of the three options (first fringes were obtained between the Keck telescopes in 2005). However, two considerations disfavor it for a link an order of magnitude longer than the 300 m OHANA demonstration. First, propagation loss: even the best-case fluoride-glass (ZBLAN) fibers, whose intrinsic K-band attenuation approaches the theoretical minimum only when manufactured in microgravity to suppress crystallization, exhibit realistic terrestrial attenuations of order 10–100 dB/km. Over the ≈3 km of fiber that would be needed to follow the sloped terrain between VISTA and the VLTI, this translates into 30–300 dB of additional loss, several to tens of magnitudes, which is incompatible with faint-target operation. Second, field transport: a single-mode fiber intrinsically filters out all spatial information beyond the fundamental mode, which is well matched to OHANA's point-source interferometric use case but would require a dense bundle of fibers, one per spatial sample, to transport an extended field or a pupil image, multiplying the cost and alignment complexity accordingly.

Evacuated light pipes, as studied for the Planet Formation Imager (PFI) design [11], avoid both issues, no turbulence, no propagation loss, full spectral coverage from the visible to the mid-infrared, but require a rigid, straight, and airtight conduit whose civil-engineering cost scales unfavorably with the sloped, kilometer-scale VISTA–VLTI terrain, and the concept remains at the design-study stage.

Free-space propagation with active AO pre-compensation, by contrast, imposes no rigid line-of-sight infrastructure beyond the two end telescopes, supports the full spectral range, and is directly compatible with a full-aperture, spatially resolved field, at the cost of requiring the turbulence along the path to be actively corrected. This is the approach retained for the VISTA–VLTI concept, and its AO dimensioning is the subject of Section 3.

### *2.4 Turbulence along horizontal propagation paths*

Atmospheric turbulence over horizontal paths differs fundamentally from the vertical columns sampled by astronomical AO. The turbulence-integrated Fried parameter $r_0$ depends on the path-weighted integral of the refractive-index structure constant $C^2_n(h)$:

$$r_0 = [0.423\, k^2 \sec(\zeta) \int C^2_n(h)\, dh]^{-3/5}$$

For a near-horizontal path, the beam samples primarily the ground layer, where $C^2_n$ is highest. At Paranal, low-altitude turbulence contributes the dominant fraction of the total seeing, and road-induced or thermal hotspots may create localized zones of strong turbulence. The hotspot distribution along the VISTA–VLTI line of sight was mapped qualitatively by Pichon et al. [6], identifying three risk zones: a low-altitude section near the VLTI, a road-crossing zone near mid-baseline, and a combined low-altitude/road zone near VISTA (Fig. 4, top). Quantitative $C^2_n$ measurements in these zones are essential to validate the AO design of Section 3.

Reference experiments over comparable horizontal paths include the FEEDELIO campaigns conducted by ONERA, which deployed AO correction up to the 7th radial order using an 8×8 sub-aperture Shack–Hartmann sensor sampled at up to 1.5 kHz, over a 13 km slant path in the Canary Islands (Teide), at a link altitude between 2400 and 3500 m [7]. This demonstration confirmed that strong AO pre-compensation of a horizontal FSO beam is feasible, but also noted that the telecom use case it addressed mostly considers high-flux, on-axis beam transmission, whereas the KBI application additionally requires a several-arcsecond field of view and works with comparatively faint targets, a distinction developed further in Section 3.3–3.4. The VISTA–VLTI path is significantly shorter (1.4 km), with the beam entirely confined within the observatory site boundary, suggesting more controlled conditions, but direct measurement is nonetheless required.

## 3. AO SYSTEM DIMENSIONING FOR THE HORIZONTAL LINK

Table 2 summarizes the top-level requirements adopted for the VISTA–VLTI AO pre-compensation system, derived from the KBI Galactic Center fringe-tracking use case and from the practical constraint of operating with realistically faint natural guide stars.

*Table 2. Top-level requirements for the VISTA–VLTI AO system.*

| Requirement | Specification | Rationale |
|---|---|---|
| Telescope size | 2 m ≤ D ≤ 8 m | D ≥ 2 m ensures sensitivity; D ≤ 8 m is the budget-driven upper limit |
| Field of view (FoV) | 4″ | Galactic Center fringe-tracking requirement |
| Limiting magnitude | H-band ≥ 10.5 | Ensures observation of faint targets |

Two (or three) architectural options were considered for splitting the correction between the on-sky (vertical) and horizontal (FSO relay) turbulence: (A) two fully independent AO loops, one operating on the vertical path at the focus of the 4 m telescope, and a second, dedicated loop correcting only the horizontal path at the VLTI platform; or (B) a single, unified multi-path loop with one AO system located at the VLTI platform correcting both paths simultaneously. Hybrid combinations (C), with partial low-order correction performed at the 4 m telescope, are also possible. The analysis below focuses on option A, sizing the dedicated horizontal-path loop independently.

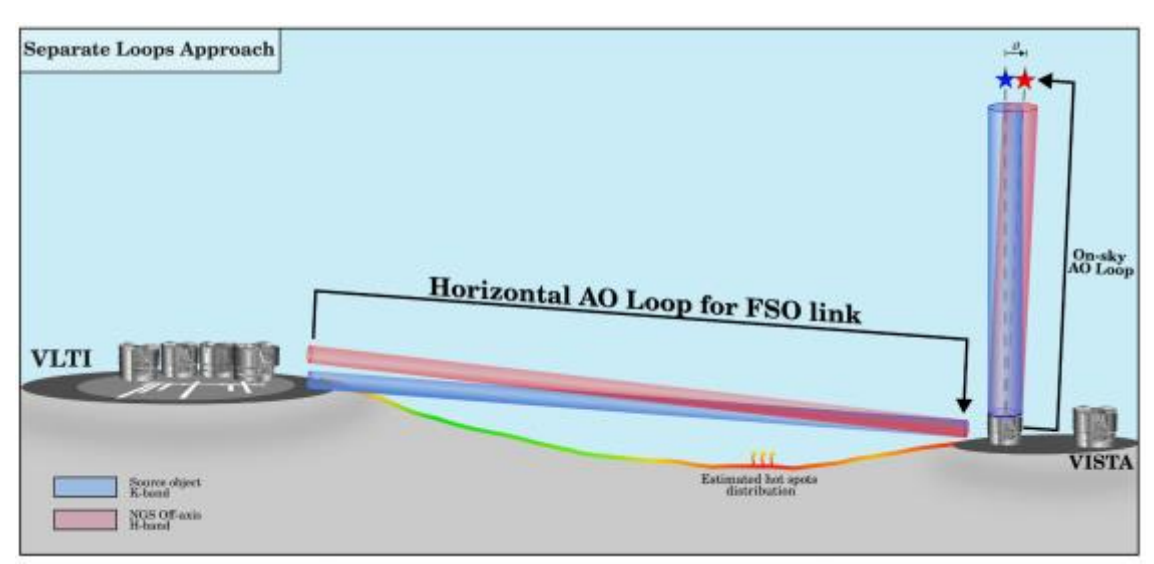

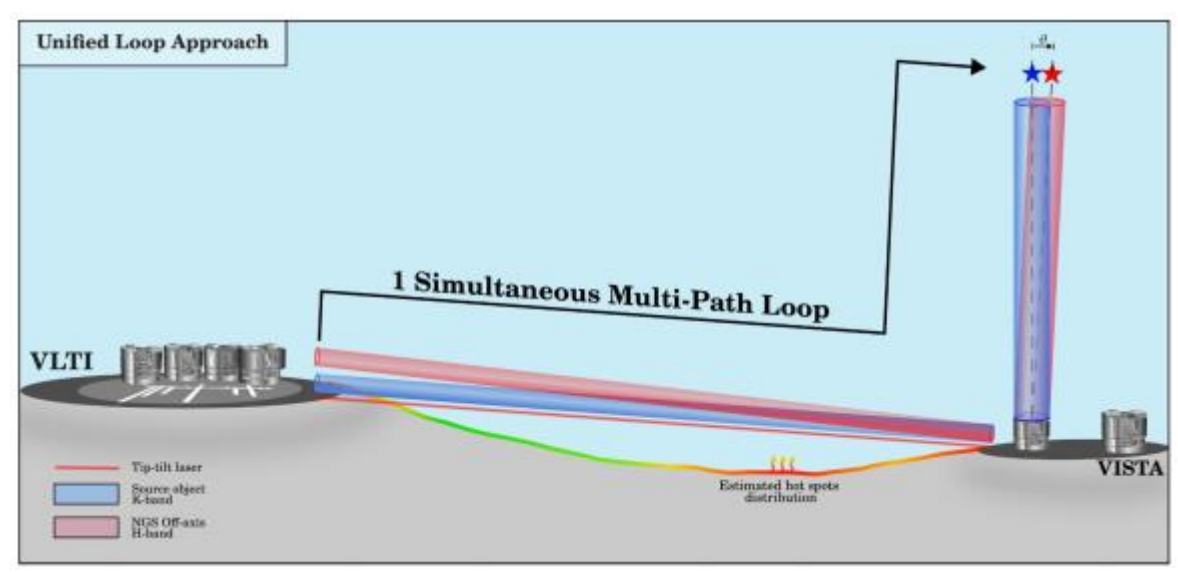

*Figure 4. The two potential implementation of the AO loops. Left would be two fully independent loops, while right would consider a single AO loop to correct both the vertical and horizontal path.*

### *3.1 Fitting error and deformable-mirror order*

The horizontal-path AO loop operates on the 0.4–0.5 m relay beam (Section 2.2) rather than on the full 4 m science aperture. For a deformable mirror with d/D actuators across the beam diameter D = 0.4 m, the fitting-error variance follows the classical relation $\sigma^2_{fit} = 0.23\,(d/r_0)^{5/3}$. Figure 5 shows the resulting wavefront-error and H-band Strehl ratio as a function of $r_0$ (referenced to 0.5 μm) for three DM configurations (5×5, 10×10, and 20×20 actuators).

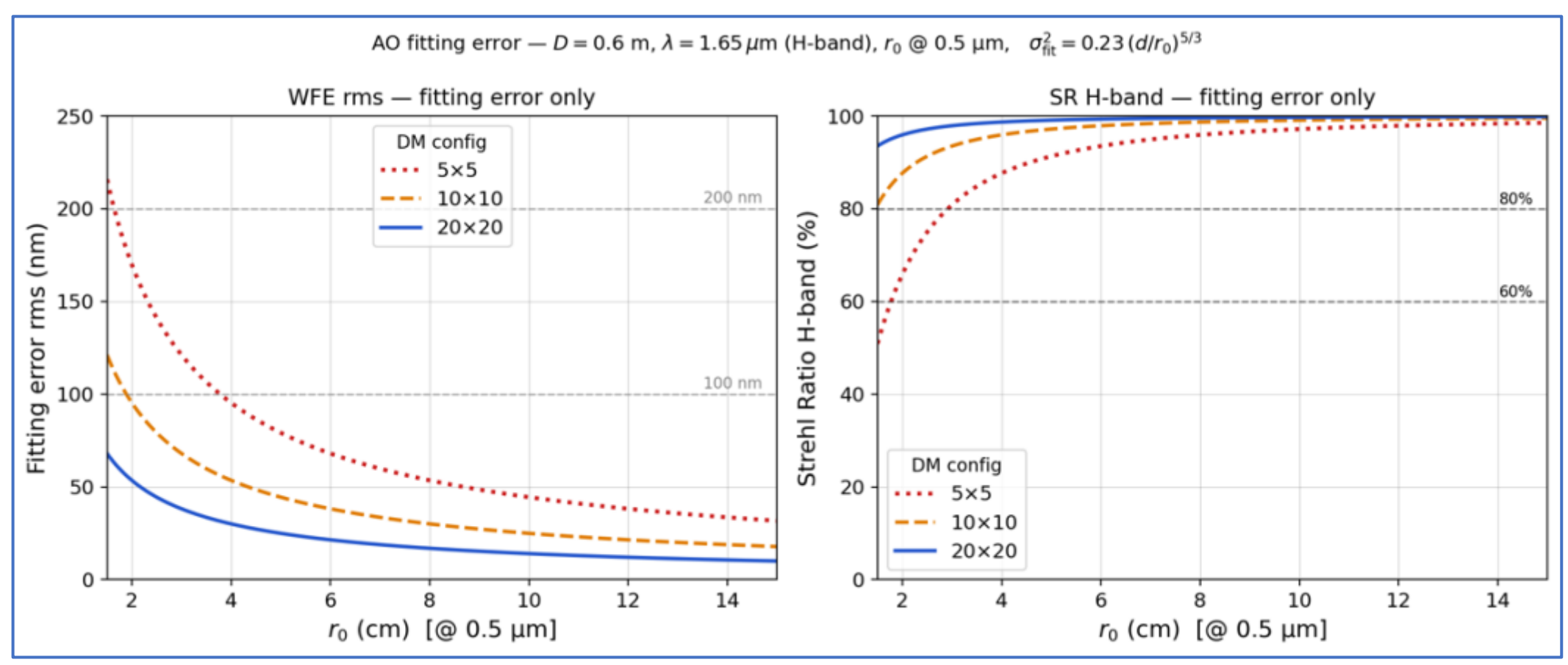

*Figure 5. AO fitting error only (D = 0.4 m, λ = 1.65 µm). Left: residual wavefront error vs. r₀. Right: corresponding H-band Strehl ratio for three DM configurations. A 10×10 actuator DM keeps the fitting-error-limited Strehl above ≈80% down to r₀ ≈ 5 cm.*

A 10×10-actuator DM appears sufficient to keep the fitting-error-limited Strehl ratio above the 60–80% range for $r_0$ down to about 5 cm, which is representative of decent-to-average seeing at Paranal. This fitting-error analysis alone, however, does not account for the photon noise of the wavefront sensor, considered next.

### *3.2 Photon budget and guide-star sensitivity*

Table 2 sets the limiting magnitude requirement at H ≥ 10.5. Figure 6 shows the number of photons per sub-aperture per frame (at a 1 kHz sampling rate) as a function of guide-star H magnitude and Shack–Hartmann sub-aperture count, together with a nominal 50-photon SNR threshold.

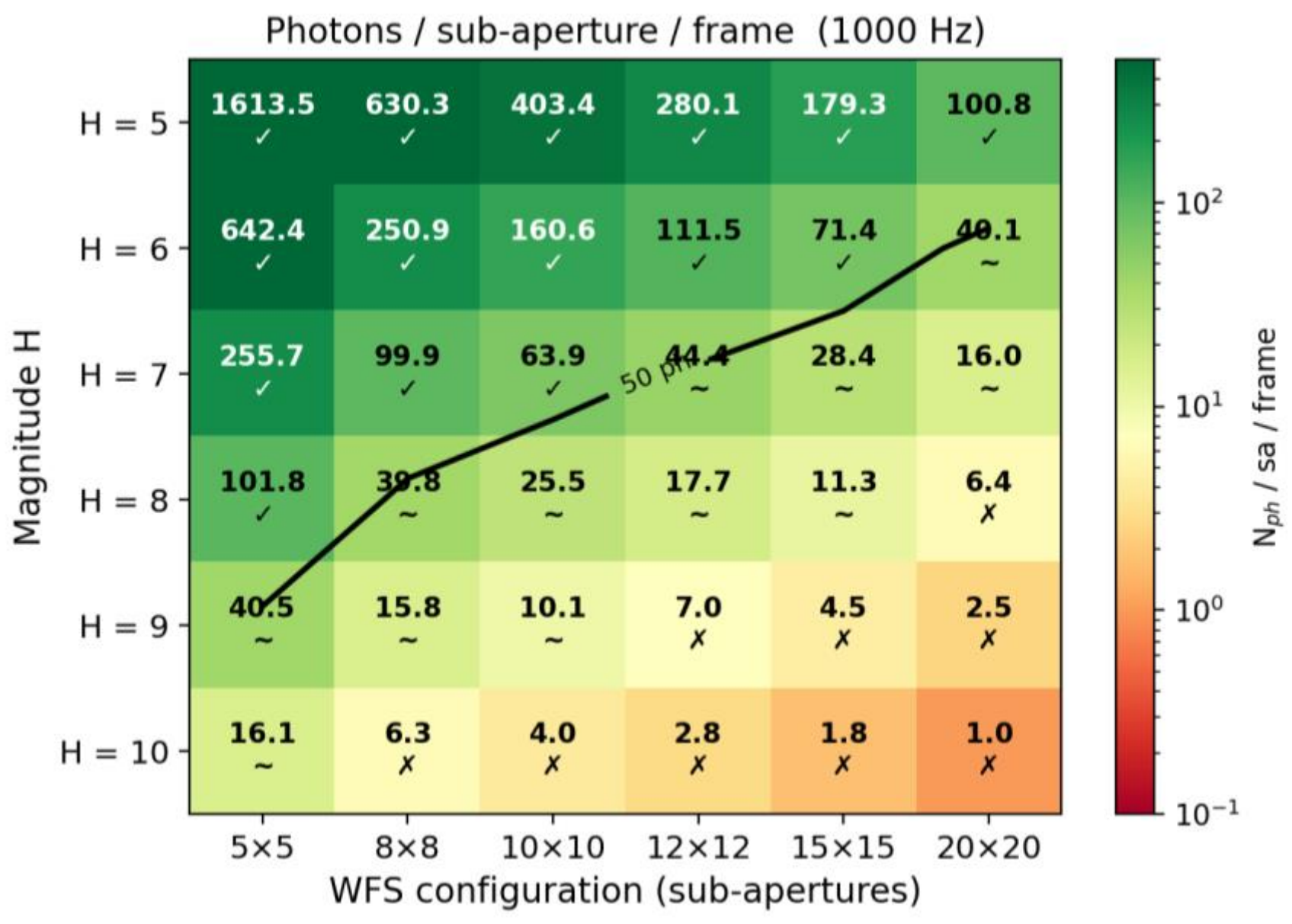

*Figure 6. Photon flux per sub-aperture per frame (1 kHz) vs. guide-star H magnitude and WFS sub-aperture configuration. The solid line marks ≈50 photons/sub-aperture/frame, a practical SNR threshold. At the H = 10.5 requirement, only coarse (5×5) configurations remain photon-sufficient.*

At H = 10.5, a 10×10 configuration, the configuration favored by the fitting-error analysis of Section 3.1, collects only about 4 photons per sub-aperture per frame, well below the noise-limited regime; only a coarse 5×5 configuration remains marginally acceptable (≈16 photons/sub-aperture/frame at H = 10). This exposes a direct tension between the actuator count needed to reach a good fitting-error-limited Strehl ratio and the actuator count that a natural guide star at the required limiting magnitude can adequately sense. The resolution adopted is a hybrid sensing scheme: a bright, dedicated laser source transmitted between VISTA and the VLTI provides a high-flux beacon for the high-order wavefront sensing (removing the fitting-error/photon-noise trade-off), while the natural guide star is used only for low-order modes (tip-tilt, focus) to track the true position and residual pointing of the astronomical source.

### *3.3 Anisoplanatism and field of view*

The Galactic Center fringe-tracking use case requires a 4″ field of view (Table 2). Because the correcting DM/WFS pair is conjugated to a single plane, the achievable field is set by anisoplanatism, whose severity depends not only on the integrated $r_0$ but critically on the distribution of $C^2_n$ along the 1.4 km path relative to that conjugate plane, information that, absent a direct measurement, is presently unknown. Three qualitative $C^2_n$ profiles were considered to bracket the possible outcomes: "ground-layer dominated" (turbulence concentrated near the telescope ends), "high-layer dominated" (turbulence concentrated away from the conjugate plane, near mid-path), and "a bit everywhere" (a spatially distributed profile) (Fig. 7).

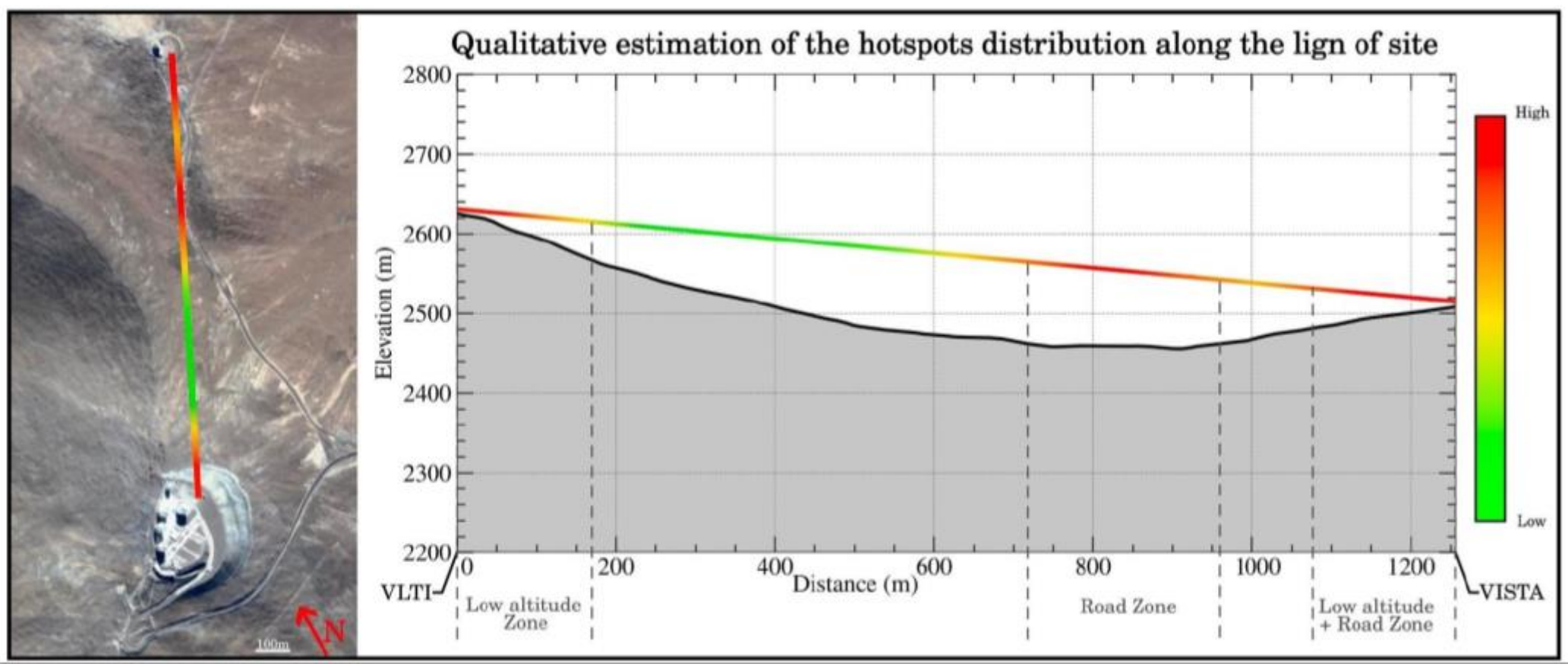

*Figure 7. Qualitative hotspot distribution along the VISTA–VLTI line of sight, and the three dummy $C^2_n(z)$ profiles considered in the absence of a direct turbulence-profile measurement.*

Figure 8 shows the resulting H-band Strehl ratio as a function of field angle for each of the three profiles, at four values of $r_0$ (15, 10, 5, and 3 cm), assuming a 12×12 actuator DM. The horizontal axis is expressed both as the physical beam angle at the 0.4 m relay aperture and as the corresponding on-sky angle after the afocal demagnification of the science beam train.

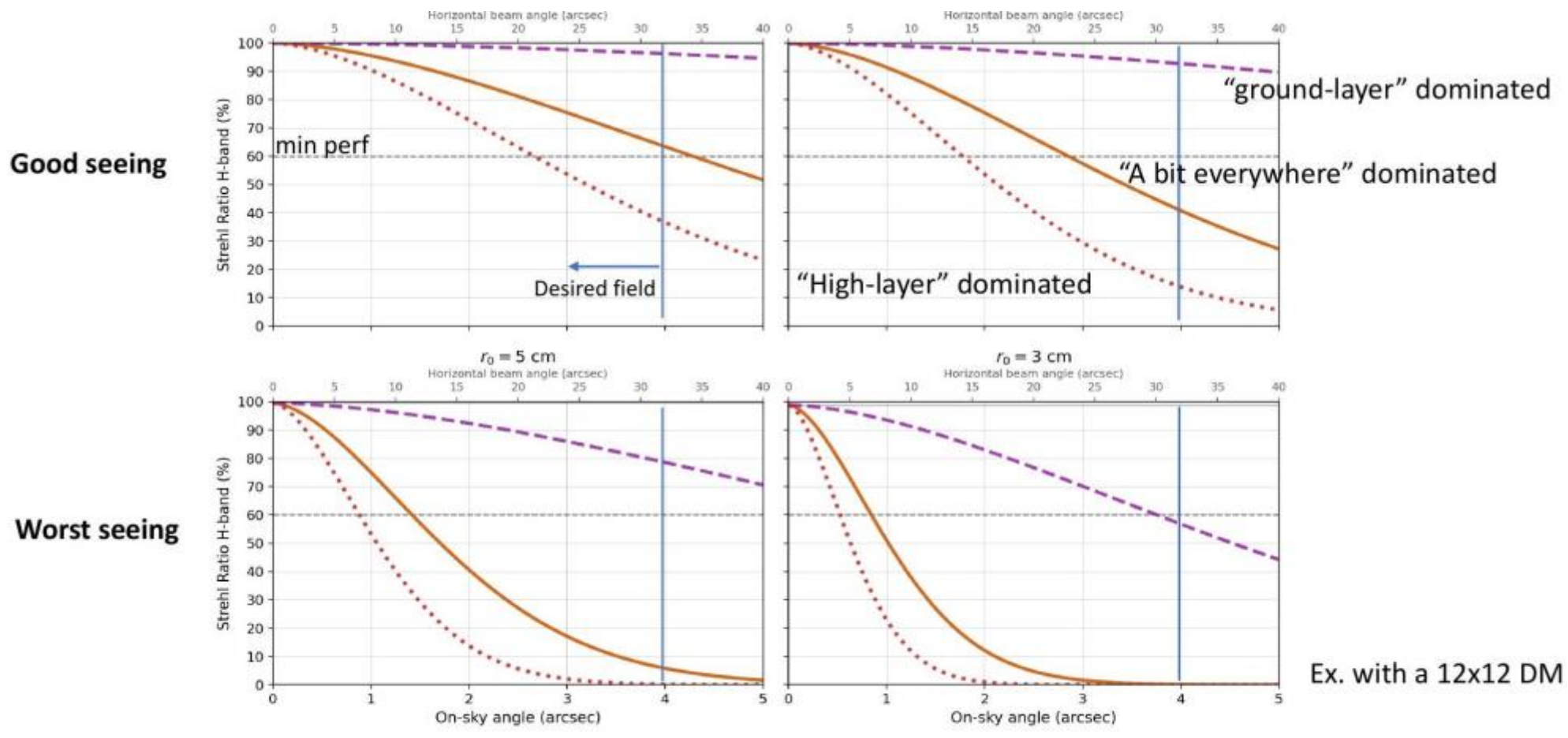

*Figure 8. H-band Strehl ratio vs. field angle for the three $C^2_n$ profiles of Fig. 7, at four values of $r_0$ (12×12 DM). The desired 4" field and a 60% minimum-performance threshold are marked. Green/red shading indicates whether the requirement is met.*

The outcome is strongly profile-dependent. Under good seeing, a "ground-layer dominated" profile keeps the Strehl ratio above the 60% threshold across the full desired field, while a "high-layer dominated" profile fails the requirement even in good seeing, and the "a bit everywhere" case sits in between. Under poor seeing ($r_0$ = 3–5 cm), only the ground-layer-dominated case comes close to meeting the field requirement. Because the true along-path $C^2_n$ distribution at Paranal for this specific line of sight is not known, this result cannot presently be used to conclude whether a single-conjugate AO system is sufficient, or whether an MCAO-like correction would be required, which is precisely the missing input that the turbulence-characterization campaign of Sections 4–5 is designed to provide.

### *3.4 Scintillation*

Beyond phase perturbations, propagation through a 1.4 km horizontal path also imprints amplitude (scintillation) fluctuations on the beam. The strength of these fluctuations is quantified by the Rytov variance, $\sigma^2_R \propto C^2_n k^{7/6} L^{11/6}$, which must remain below unity, and preferably below ≈0.1, for the weak-fluctuation propagation theory to remain valid. From first computations, the analytic Rytov parameter is estimated at $\sigma^2_R \approx 0.06$, remaining below the 0.1 threshold across all realistic Paranal conditions.

## 4. HORIZONTAL TURBULENCE MEASUREMENT: EXPERIMENTAL CONCEPT

### 4.1 Overall architecture

The turbulence characterization campaign explores two complementary implementations for generating and sampling the calibrated beacons along the 1.4 km baseline, of increasing complexity.

The primary diagnostic instrument is a Shack–Hartmann wavefront sensor (SHWFS) operating in wide-field mode, observing an array of beacons. In this configuration, each sub-aperture of the lenslet array records the image displacement of multiple beacons placed at known angular separations in the field. The differential displacement between beacons as a function of sub-aperture position and beacon separation provides the raw observable for tomographic reconstruction. The reconstruction technique follows the SLODAR family of methods, in which the cross-covariance of slope measurements for different beacon pairs peaks at a position in the covariance plane that encodes the altitude of the contributing turbulent layer [8]. For a horizontal path, the altitude is replaced by the along-beam distance, and the reconstruction yields the $C^2_n(z)$ profile as a function of distance z from the transmitter, directly analogous to the SLODAR technique applied to vertical profiles, here adapted to the horizontal geometry.

The SHWFS output will provide: (1) the integrated $r_0$ along the full path, (2) the longitudinal $C^2_n(z)$ profile with a depth resolution determined by the beacon angular separation and the sub-aperture diameter, and (3) the outer scale $L_0$ and the temporal evolution of each parameter. If some scintillation is present, we could also implement a CO-SLIDAR technique, which jointly uses the slopes and scintillation indices from a double source recorded with a SHWFS [12].

We have been exploring two potential implementations described below.

**Option 1 — single telescope and a fixed source grid.** In its simplest form, a single compact telescope (30–50 cm) is installed at one end of the baseline and observes a grid of calibrated sources (LEDs or fiber tips) mounted on a rigid bench at a fixed distance in front of it. Each source subtends a small but non-zero angle at the telescope, so a SHWFS lenslet array sees a set of beacons at known angular separations, and the same SLODAR-type cross-covariance analysis described in Section 5.1 can be applied directly.

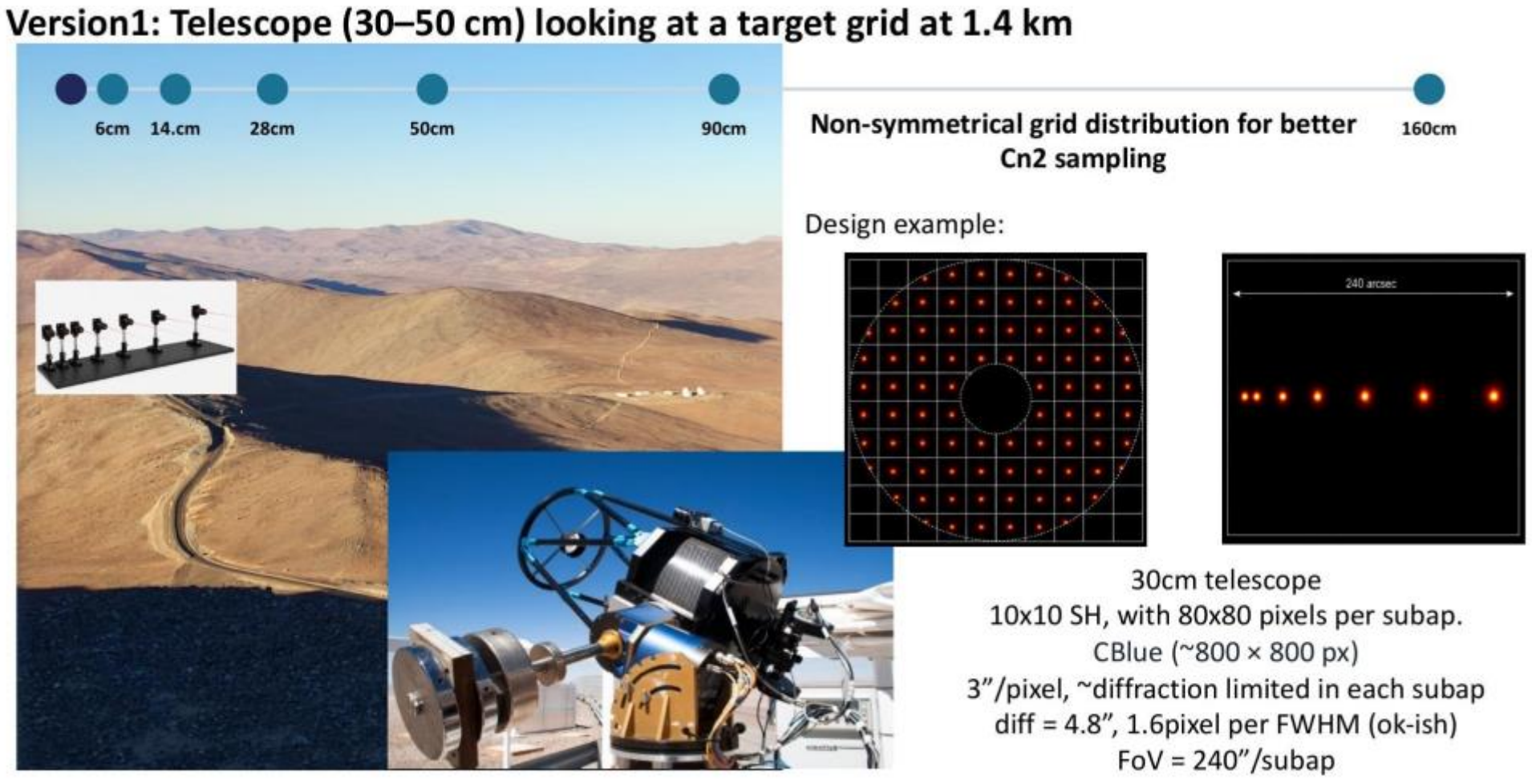

*Figure 9. Two-telescope facing configuration: a first telescope (the "projector") places calibrated sources (LEDs or fibers) at its focus, so that each source emerges collimated — a virtual star at infinity — sampling the full 1.4 km path, in the manner of a classical horizontal SLODAR experiment.*

The key design driver for this option is the choice of source spacing. Widely separated sources (up to ≈160 cm on the bench) finely sample the turbulence near the receiving telescope but under-sample the turbulence located close to the source end of the path; conversely, closely spaced sources (down to ≈6 cm) preferentially sample the turbulence near the sources themselves, at the cost of missing structure near the receiver. In practice, the maximum usable source separation is set by the field of view of a single SHWFS sub-aperture, while the minimum separation is set by the diffraction limit of the spots on the detector. A non-symmetrical grid (6, 14, 28, 50, 90, and 160 cm spacings in the current design) is therefore adopted to balance the sampling of the near- and far-end turbulence within a single, simple setup. A representative design uses a 30 cm telescope with a 10×10 sub-aperture SHWFS (≈80×80 pixels per sub-aperture) on a CBlue camera, at a plate scale of 3″/pixel; this yields a diffraction-limited spot size of ≈4.8″ (≈1.6 pixels per FWHM) and a field of view of 240″ per sub-aperture.

This single-telescope configuration is straightforward to deploy and calibrate, but it is intrinsically anchored to one end of the path: it can be optimized to sample either the near-receiver or the near-source turbulence well, but not both simultaneously over the full 1.4 km.

**Option 2 — two telescopes facing each other.** This upgrade removes this limitation by deploying a second, identical compact telescope at the opposite end of the baseline. Rather than illuminating a physical grid at a fixed short distance, the far telescope is operated as a "projector": the calibrated sources are placed at its own focus, so that each one emerges collimated, a virtual star at infinity, exactly as in a classical astronomical SLODAR set-up, and the resulting beacon grid samples the atmosphere along the *entire* 1.4 km path rather than only its immediate vicinity. This configuration also makes it straightforward to provide several simultaneous baselines and keeps each virtual-star pair comfortably within the field of view of a single SHWFS sub-aperture, since the angular separations are now set by the focal-plane source positions rather than by a physical bench.

Each of the two units therefore hosts both a calibrated light source (in emission/projector mode) and a turbulence sensor (in reception mode), enabling reciprocal beam propagation and, in principle, simultaneous two-way turbulence measurement. The transmitter chain at each site consists of a stabilized near-infrared laser source (H or K band) or LED array feeding the projector telescope; the receiver chain feeds the incoming beam into one or both sensing modalities described in Section 5. The experiment is designed to be operated in a campaign mode, collecting data over a range of weather conditions (wind speed, ambient temperature, time of night) to build a statistical database of horizontal turbulence at Paranal.

The choice between Option 1 and Option 2 directly affects the achievable depth coverage: only the facing-telescope configuration (Option 2) samples the full 1.4 km path with a single beacon grid, and is therefore best suited to resolving the anisoplanatism uncertainty identified in Section 3.3, the reconstructed $C^2_n(z)$ profile will indicate whether the true turbulence distribution is closer to the "ground-layer" or "high-layer" dominated case of Fig. 7–8, and hence whether a single-conjugate AO correction is adequate for the required 4″ field.

### *4.2 Event-based cameras for high-speed turbulence signatures*

A complementary sensing modality will exploit event-based cameras (also called neuromorphic or dynamic vision sensors). Unlike conventional frame-based detectors, event cameras respond to changes in log-intensity at each pixel independently and asynchronously, with a temporal resolution of order 1 μs and a dynamic range exceeding 120 dB [9]. Each pixel fires an event when its log-intensity changes by a threshold amount, producing a sparse stream of (x, y, t, polarity) events rather than dense frames.

This architecture offers two key advantages for turbulence characterization. First, the microsecond temporal resolution resolves scintillation fluctuations well above the Greenwood frequency (typically 10–1000 Hz for strong horizontal turbulence), allowing the power spectral density of intensity fluctuations to be measured over several decades of frequency, directly probing the amplitude fluctuations discussed in Section 3.4. Second, the spatial read-out is truly independent per pixel, so anisotropic turbulence signatures, such as elongated scintillation patterns produced by wind shear in a thin layer, can be detected and characterized with a statistical fidelity that conventional cameras cannot achieve without prohibitively high frame rates and data volumes.

The event camera will observe the pupil image of the collimated beam after propagation across the 1.4 km path. Temporal correlation of the event stream across the pupil will provide the spatial covariance of log-amplitude fluctuations, which is directly related to the turbulence spectrum through the classical scintillation theory. The complementarity with the SHWFS is important: the SHWFS measures phase perturbations, while scintillation probes amplitude perturbations arising from propagation through turbulence at finite distance. Together, they provide a more complete characterization of the turbulence, and a direct empirical check of the weak-scintillation assumption made in Section 3.4.

The event-camera turbulence-sensing methodology has been demonstrated experimentally in a series of field campaigns conducted at the SEETRUE Optical Ground Station, located at the Curauma Campus of Pontificia Universidad Católica de Valparaíso, Chile [13]. The receiver system used two Celestron EdgeHD 14-inch telescopes, each equipped with a SilkyEvCam neuromorphic sensor, observing passive high-contrast checkerboard targets at the far end of horizontal atmospheric paths of approximately 200 m and 300 m. The two independent channels sampled nearly the same atmospheric volume, providing cross-validation of the retrieved turbulence quantities.

The measurement concept exploits the asynchronous pixel response of the event camera. For each temporal integration window $\Delta t$, the event stream is converted into a relative irradiance sequence by signed event accumulation. From the reconstructed irradiance, the pixel-wise scintillation index $\sigma^2_i(r)$ is computed as the normalized variance of intensity fluctuations. Assuming a spherical-wave propagation model, the path-averaged refractive-index structure constant is then retrieved as $C^2_n \approx 4.0\, \sigma^2_i\, k^{-7/6}\, L^{-11/6}$, where L is the path length, $k = 2\pi/\lambda$ is the optical wavenumber, and $\lambda$ is the effective wavelength.

The SEETRUE campaigns demonstrated that event cameras contain recoverable image-plane scintillation information sufficient to estimate $C^2_n$ over near-ground horizontal paths. The event-camera estimates were compared with a sonic anemometer reference and showed agreement at the level of hourly statistics under well-illuminated conditions. The main limitations identified were mechanical vibration from wind loading, which can generate spurious events unrelated to atmospheric scintillation, and insufficient scene contrast under low-light conditions. These findings directly motivate two design choices in the VISTA–VLTI campaign: improved mechanical stabilization of the receiver optics, and the use of controlled near-infrared LED arrays (850 nm) as calibrated active beacons to ensure a stable and repeatable signal under nighttime conditions.

## 5. CONCLUSION

We have presented the concept and design of a dedicated turbulence characterization experiment for the VISTA–VLTI 1.4 km baseline at Paranal, together with a preliminary AO dimensioning study for the free-space optical link itself. Free-space propagation with active AO correction was retained over optical fiber and evacuated-pipe alternatives, primarily because of the prohibitive propagation losses of long fluoride-glass fibers in the K band and their inability to transport spatially resolved field information, and because of the civil-engineering cost of a kilometer-scale evacuated pipe on sloped terrain. AO dimensioning indicates that a moderate-order ($\approx 10\times10$ actuator) correction can, in principle, satisfy the fitting-error budget, but that photon-starved natural guide stars require a hybrid sensing scheme combining a bright artificial beacon with low-order natural-guide-star tracking, and that the achievable field of view is highly sensitive to the largely unknown along-path distribution of $C^2_n$. This last point is the central open question that the proposed experiment is designed to resolve.

The experiment will deploy two complementary sensing modalities, a wide-field Shack–Hartmann sensor for tomographic turbulence profiling, and an event-based camera for high-speed scintillation characterization, to provide a comprehensive statistical database of horizontal atmospheric turbulence over this path. This database constitutes the critical missing input for finalizing the design of the AO pre-compensation system that will enable a free-space optical beam link between the VISTA and VLTI sites.

The broader context is the growing scientific and technical case for Kilometer Baseline Interferometry in the optical/near-infrared. The science enabled by 10–100 μas resolution, exoplanet surface imaging, circumplanetary disk spectro-astrometry, black-hole physics from galactic to cosmological scales, motivates a sustained development effort, of which the VISTA–VLTI proof-of-concept is the natural first step. The turbulence campaign described here will begin in mid-2026, with results expected to be available in time to inform the design reviews for the extended VLTI outrigger telescope.

## ACKNOWLEDGMENTS

This work benefited from the support of the French National Research Agency (ANR) through grants WOLF (ANR-18-CE31-0018), APPLY (ANR-19-CE31-0011), and LabEx FOCUS (ANR-11-LABX-0013); the Programme Investissement Avenir F-CELT (ANR-21-ESRE-0008); the Action Spécifique Haute Résolution Angulaire (ASHRA) of CNRS/INSU co-funded by CNES; the Région Sud; and the French government under the France 2030 investment plan as part of the Initiative d'Excellence d'Aix-Marseille Université – A*MIDEX (AMX-22-RE-AB-151).